\documentstyle[preprint,prd,aps,floats,epsf]{revtex}
\tighten	
\def\defn#1{``#1''}

\def\eg{e.g.\hbox{}}
\def\ie{i.e.\hbox{}}

\def\eqref#1{\hbox{(\ref{#1})}}
\def\constant{{\text{constant}}}
\def\del{\nabla}
\def\tfrac#1#2{{\textstyle\frac{#1}{#2}}}
\def\thalf{\tfrac{1}{2}}
\def\tr{\mathop{\rm tr}\nolimits}
\def\E{\bbox{E}}
\def\B{\bbox{B}}
\begin{document}
\title{Symmetry without Symmetry: Numerical Simulation of \\ Axisymmetric 
       Systems using Cartesian Grids}
\author{\hbox{Miguel Alcubierre$^{(1)}$},
	\hbox{Steven Brandt$^{(2)}$},
	\hbox{Bernd Br\"{u}gmann$^{(1)}$},
	\hbox{Daniel Holz$^{(1)}$},
	\hbox{Edward Seidel$^{(1,3)}$},
	\hbox{Ryoji Takahashi$^{(1)}$},
	and \hbox{Jonathan Thornburg$^{(4)}$}}
\address{\mbox{}\\
         $^{(1)}$
	 Max-Planck-Institut f\"ur Gravitationsphysik, 
         Albert-Einstein-Institut,
	 Am M\"uhlenberg~1, D-14476 Golm, Germany			\\
	 $^{(2)}$
	 Department of Physics,
	 Pennsylvania State University, State College, PA 16802, USA	\\
	 $^{(3)}$
	 National Center for Supercomputing Applications,
	 Beckman Institute, 405~N.~Mathews Ave., Urbana, IL 61801, USA	\\
	 $^{(4)}$
	 Institut f\"{u}r Theoretische Physik, Universit\"{a}t Wien,
	 Boltzmanngasse~5, A-1090 Wien, Austria}
\date{\today; AEI-1999-14}
\maketitle
\begin{abstract}
We present a new technique for the numerical simulation of
axisymmetric systems. This technique avoids the coordinate
singularities which often arise when cylindrical or polar-spherical
coordinate finite difference grids are used, particularly in
simulating tensor partial differential equations like those of $3+1$
numerical relativity.  For a system axisymmetric about the $z$~axis,
the basic idea is to use a 3-dimensional {\em Cartesian\/} $(x,y,z)$
coordinate grid which covers (say) the $y=0$ plane, but is only one
finite-difference-molecule--width thick in the $y$~direction.  The
field variables in the central $y=0$ grid plane can be updated using
normal $(x,y,z)$--coordinate finite differencing, while those in the
$y \neq 0$ grid planes can be computed from those in the central plane
by using the axisymmetry assumption and interpolation.  We demonstrate
the effectiveness of the approach on a set of fully nonlinear test
computations in $3+1$ numerical general relativity, involving both
black holes and collapsing gravitational waves.
\end{abstract}
\draft


\section{Introduction}
\label{sec:intro}

Finite difference numerical simulations of axisymmetric systems are
most often, and most naturally, carried out in coordinate systems
adapted to the symmetry of the underlying problem, \eg{} polar
spherical $(r,\theta,\phi)$ or cylindrical $(\rho,z,\phi)$
coordinates.  However, the use of such coordinate systems brings with
it delicate problems in finite differencing near the $z$~axis,
particularly when {\em tensor\/} time-evolution partial differential
equations (PDEs) are considered. Depending on the problem, it is often
very difficult to obtain fully stable numerical evolutions near the axis,
and for some problems it is difficult to even accurately discretize
the equations there. Here we consider the problem for the Einstein
equations of general relativity in axisymmetry, although our approach
should be useful for other systems of PDEs (\eg{}~the Navier-Stokes
equations), and also for the case of spherical symmetry.

There are several different types of $z$~axis difficulties, which
depending on the physical system may occur singly or in combination.
The simplest problem is that physically nonsingular terms in the
equations may have indeterminate $0/0$ forms along the $z$~axis.
Fortunately, such terms are generally easy to regularize by applying
L'Hopital's rule.  For example, in polar spherical coordinates
$(r,\theta,\phi)$, the flat-space Laplacian operator includes the term
\begin{equation}
	\frac{1}{r^2 \sin\theta} \,
   	   \partial_\theta \Big( \sin\theta \, \partial_\theta \Big)
					\,\, \text{.}	
						\label{eqn:Laplacian-term}
\end{equation}
Assuming all fields to be smooth on the $z$~axis and applying
L'Hopital's rule, the $\theta \to 0$ limit of this term is easily
seen to be
\begin{equation}
	\frac{2}{r^2} \, \partial_{\theta\theta}
					\,\, \text{.}	
					\label{eqn:Laplacian-term-z-axis}
\end{equation}

However, even here there may be a problem with finite differencing:
Although the $\theta \to 0$ limit of the term~\eqref{eqn:Laplacian-term}
is precisely~\eqref{eqn:Laplacian-term-z-axis}, when we finite difference
these terms the limiting relationship generally only holds in the limit
where the grid spacing $\Delta\theta \to 0$.  For any given (nonzero)
$\Delta\theta$, the numerically computed values of~\eqref{eqn:Laplacian-term}
will have a $\theta \to 0$ limit which will in general differ somewhat
from the numerical values of~\eqref{eqn:Laplacian-term-z-axis}.
This difference may give rise to finite differencing instabilities
near the axis.

A more serious $z$~axis problem is that of {$\infty-\infty$ 
cancellations}, where a physically nonsingular quantity is computed as 
the sum of many terms, two or more of which are individually singular 
on the $z$~axis.  For example, again using polar spherical coordinates, 
$(r,\theta,\phi)$, in the so-called ``$3+1$'' Einstein equations 
(\cite{Arnowitt62}; see, for example, \cite{York79,York83} for general 
reviews), the 3-Ricci tensor component $R_{\theta\theta}$ is 
generically $O(1)$ on the $z$~axis, but it is computed as the sum of, 
among (many) other terms, $- \thalf g^{\phi\phi} 
\partial_{\theta\theta} g_{\phi\phi}$, which is generically 
$O(1/\theta^2)$ near the $z$~axis.  Although not impossible, 
regularizing terms of this nature is very difficult, requiring a 
detailed analysis of the generic behavior of the entire system of PDEs 
under consideration (for our example, the coupled $3+1$ evolution of 
Einstein's equations, with constraint and coordinate equations) near 
the $z$~axis.

Collectively these problems can be very severe, crippling attempts to
evolve systems with such symmetries (see,
\eg{}~\cite{Smarr75,Bernstein93b,Thornburg93,Anninos94a}).  Many
solutions to the problems brought on by special coordinate systems
have been attempted, including various regularization
procedures~\cite{Evans86,Stark91,Seidel88a,Seidel90c}, Taylor
expansions~\cite{Seidel88a}, the use of nonsingular-basis tensor
components~\cite{Thornburg93}, spectral methods~\cite{Bonazzola89},
and special coordinate conditions used to eliminate troublesome
terms~\cite{Smarr75,Bernstein93a,Bernstein93b}. However, these
methods tend to be complicated and not particularly robust.

On the other hand, 3D~Cartesian coordinates contain no coordinate
pathologies, and all terms in the equations governing the evolution of
functions are typically completely regular, even at special points
such as the origin or along a symmetry axis~\cite{Bardeen83}. 
Normally, however, if one treats a spherical or axisymmetric system in 3D
Cartesian coordinates one loses the ability to ignore irrelevant
dimensions.  For example, a spherically symmetric system should reduce
to a 1D problem, depending only on the radius $r$, while an
axisymmetric system reduces to a 2D problem, independent of the
azimuthal coordinate $\phi$.  In full 3D~Cartesian coordinates, not
only is the true symmetry of the problem potentially disturbed by the
finite differences taken in a Cartesian coordinate system, but
dimensional reductions do not occur, and the memory requirements are
much larger, scaling as $N^{3}$, rather than as $N^{2}$ or $N$.  For
reasonable grid sizes of order hundreds of zones, these factors can
become astronomical.  In 3D numerical relativity, the problem is
exacerbated by the need to carry a large number of 3D grid functions
(typically over 100) in memory at all times.  Some savings have been
realized in cases where Cartesian coordinates are used for
intrinsically axisymmetric or spherical problems by evolving only a
single octant, reducing memory and computational requirements by a
factor of eight, but this does not change the overall scaling with
$N$.

Furthermore, while many problems are now treatable in 3D, where 
Cartesian coordinates are generally favored, testbeds for developing 
algorithms for the full 3D case are often carried out in lower 
dimensional cases, where the problems are simpler and where limiting 
solutions are known. However, working in special coordinate systems 
for lower dimensional test problems introduces difficulties specific 
to the coordinate system, and frequently techniques developed in 
special coordinate system do not carry over to the 3D~Cartesian code.  
What is needed is a lower dimensional testbed that retains the same 
essential features present in the generic 3D case, at both the 
physical and computational level.

In this paper we describe a scheme which borrows from the singularity-free
nature of full 3D~Cartesian coordinates, and allows the treatment of
axisymmetric
and spherically symmetric problems without the memory constraints of full
3D~Cartesian coordinates.  Take the case of axisymmetry:  The essential
trick is to realize that in 3D~Cartesian coordinates, an axisymmetric
system can be computed in, say, the $x$-$z$ ($y=0$) plane alone.  The
$x$-$z$ system can be rotated about the $z$~axis to determine the solution
at any $(x,y,z)$~point at a given instant of time.  However, a
3D evolution also requires spatial derivatives in the $y$~direction,
which (for non-scalar field variables) do not necessarily all vanish,
even in the $x$-$z$ ($y=0$) plane.  Because of the axisymmetry assumption,
the solution in the $x$-$z$ ($y=0$) plane can be rotated according
to tensor transformation laws to any $y$~value, so the $y$~derivatives
of all quantities can be determined in the $x$-$z$ plane solely from
information in this same plane.  Hence, a full
3D~evolution in Cartesian coordinates can be carried out, using
{\em only\/} information from a single 2D~plane of data.  There are
different ways to achieve this, as we show in the sections below, but
the key point is that an axisymmetric or spherical system can be
evolved {\em as if it were in 3D~Cartesian coordinates\/}, and hence
without coordinate singularities and the instabilities they can
induce, and without going to the expense of a full 3D computation.

In section~\ref{sec:technique} we describe the technique in detail for
the case of axisymmetry. We show the effectiveness of the method in
applications to dynamic wave and black hole spacetimes in
section~\ref{sec:apps}, and summarize the work in
section~\ref{sec:concl}.  The method has been fully implemented under
the name \defn{Cartoon} (chosen because of its resemblance to the way
low-budget television cartoons animate a nominally 3-dimensional world
in ``$2\thalf$''~dimensions, and also as a shorthand for {\em
car}tesian {\em two}-dimensional) in the Cactus code for numerical
relativity and
astrophysics~\cite{Cactusweb,Allen99a,Bona98b,Seidel98c}.


\section{The Technique}
\label{sec:technique}

In 3D systems with {\em discrete\/} symmetries it has been common
practice to evolve just a part of the domain, and use symmetries
to provide boundary conditions.  For example, spherical, axisymmetric,
and even 3D systems have been evolved in numerical relativity in a
single octant, if the symmetries of the system allow it.  A spherical
system is reflection symmetric about all coordinate planes, and hence
it could be evolved in Cartesian coordinates in the octant defined by,
say, $x,y,z$ all non-negative. Any planar reflection symmetry can be
used to provide boundary conditions at the plane of reflection,
\ie{}~from the symmetry one can infer that certain functions are
either symmetric or antisymmetric across the coordinate planes.
This has been used in many simulations in numerical relativity
\cite{Bona98b,Anninos94c,Anninos94d,Bruegmann96,Camarda97a,
Camarda97b,Camarda97c,Allen98a,Allen98b}.

In what follows we discuss how in the case of axisymmetry one can use
the {\em continuous\/} rotational symmetry to provide boundary
conditions for a thin 3D slab. The same construction
is expected to be applicable to spherically symmetric systems.

\begin{figure}      
\epsfxsize=100mm   
\hspace{15mm}
\epsfbox{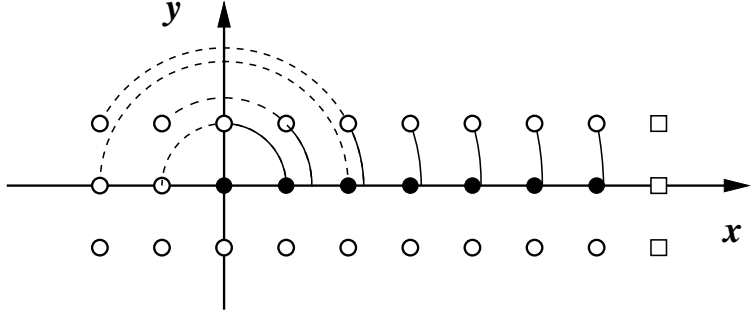}
\vspace{5mm}     
\caption{
A typical numerical grid in a constant $z$ plane with $\Delta x=\Delta
y$, no staggering, for a 3D molecule of radius 1.  The arcs show
the rotation allowed by axisymmetry to obtain various field values on
the slab boundary (circles).  High order interpolation in the
$x$~direction may require several points at $x<0$; for these points
the rotation is shown by dashed arcs.
The outer boundary at $x_{\max}$ is given as part of the problem (squares).
In the interior of the grid at $0\leq~x<x_{\max}$, $y=0$ (solid dots),
the standard 3D finite difference stencil can be used.
}
\label{fig:slab}      
\end{figure}

Consider a rectangular three dimensional volume in $R^3$, with
Cartesian coordinates $(x,y,z)$.  We assume a uniformly spaced
Cartesian grid with spacings $\Delta x$, $\Delta y$, and $\Delta z$,
and we take the finite difference molecules to have radii~$v$ in the
$x$~direction and $w$~in the $y$~direction.
(For reasons discussed in section~\ref{sec:interpolation}, we will
typically have $v > w$.)  We take the $z$~axis to be the axis of
symmetry, as shown in Fig.~\ref{fig:slab}.  In the $y$~direction, the
grid contains $2w+1$ points centered about $y = 0$.  (For example,
for standard second order centered difference $w=1$, and there are
points at $y = 0$ and $\pm \, \Delta y$.)  In the $x$~direction,
the grid starts near the axis at $x_{\min}$ and extends out to some
finite positive $x_{\max}$.  We consider the two cases
$x_{\min} = - (v-\thalf) \Delta x$ (staggered) and
$x_{\min} = - v\Delta x$ (non-staggered).

One can compute finite differences with a centered Cartesian 3D
molecule in the interior plane of the slab, \ie{}~for all grid points
at $x \geq 0$ and $y = 0$, once boundary values are specified. The
boundary values in the $z$~direction and the positive $x$~direction
are assumed to be given as part of the general problem. As outlined
above, the key point in the case of axisymmetry is that field
values at points with $y\neq 0$, and also at the points with $x < 0$,
can be obtained through a rotation and interpolation from field
values on the half plane $x\geq 0$ and $y=0$.

\subsection{Rotation of Tensors}

Let us first discuss the rotation of the tensors and non-tensors that
arise in typical numerical relativity computations. The basic formula, 
Eq.~(\ref{basicrot}) below, is (aside from a few subtleties in its derivation)
precisely what one expects for a vector in $R^3$ in Cartesian
coordinates. Consider the rotation
of a tensor field by an angle $-\phi_0$ about the $z$ axis.  
Equivalently, we can view this as keeping the field fixed, but
rotating the {\em coordinates\/} in the {\em opposite\/} direction,
\ie{} rotating them by $+ \phi_0$.  Taking this latter viewpoint,
in cylindrical coordinates $(\rho,\phi,z)$ that are adapted to the
rotation about the $z$~axis, the rotated coordinates are
\begin{equation}
	\rho' = \rho, \quad \phi' = \phi + \phi_0, \quad z' = z,
\end{equation} 
or in the Cartesian coordinates given by $\rho=\sqrt{x^2+y^2}$,
$\phi=\arctan(y/x)$,
\begin{equation}
	x' = x \cos\phi_0 - y \sin\phi_0, \quad
        y' = x \sin\phi_0 + y \cos\phi_0, \quad
        z' = z.
\end{equation}
This coordinate transformation gives rise to the linear map
\begin{equation}
   ({R(\phi_0)^i}_{j}) = \left(\frac{\partial {x'}^i}{\partial {x}^j} \right)
               = \left(\begin{array}{ccc}
                       \cos\phi_0 & -\sin\phi_0 & 0 \\
                       \sin\phi_0 & \cos\phi_0 & 0 \\
                          0       &      0     & 1 
                       \end{array}\right).
\label{rotation-coord-xform}
\end{equation}
Note that $R(\phi_0)^{-1} = R(-\phi_0)$.

Now consider arbitrary tensor fields $T$ on a three-dimensional
manifold $\Sigma$. Without referring to coordinates, a rotation (if it
exists) defines a diffeomorphism $R:\Sigma\rightarrow\Sigma$, mapping
points to points, and as such also defines a new tensor field $R^*T$
of the same contra- and covariant type as $T$. The diffeomorphism $R$
is a symmetry transformation for the tensor field $T$ if $R^*T = T$.

We assume, as is often the case in numerical relativity, that the
domain of interest is covered by a single coordinate chart. The matrix
of components of $R^*$ in the coordinate bases of the coordinate
system $(x^i) = (x,y,z)$ at a point $p$ and the coordinate system
$(x'^i) = (x', y', z')$ at the point $R(p)$ equals the Jacobian matrix
of the map $R$ between the coordinates, which is given 
in~\eqref{rotation-coord-xform},
if we assume coordinates adapted to the rotation. The components
of a tensor $T$ at $p$, for example ${T^{i_1, i_2, \ldots}}_{j_1, j_2,
\ldots}(p)$, transform according to
\begin{equation}
{(R^* T)^{i_1, i_2, \ldots}}_{j_1, j_2,\ldots} (R(p))
= {R^{i_1}}_{k_1} {R^{i_2}}_{k_2} \ldots
  {(R^{-1})^{l_1}}_{j_1} {(R^{-1})^{l_2}}_{j_2} \ldots
{T^{k_1, k_2, \ldots}}_{l_1, l_2,\ldots}(p) .
\label{tensor-coord-xform}
\end{equation} 
In the case of axisymmetry ($R^*T=T$) we therefore arrive at 
\begin{equation}
{T^{i_1, i_2, \ldots}}_{j_1, j_2,\ldots} (x,y,z)
=
{R^{i_1}}_{k_1} {R^{i_2}}_{k_2} \ldots
  {(R^{-1})^{l_1}}_{j_1} {(R^{-1})^{l_2}}_{j_2} \ldots
{T^{k_1, k_2, \ldots}}_{l_1, l_2,\ldots} (\sqrt{x^2+y^2}, 0, z),
\label{basicrot}
\end{equation} 
where $R$ is given by~\eqref{rotation-coord-xform} with $\cos\phi_0 = x/\rho$,
$\sin\phi_0 = y/\rho$, and $\rho = \sqrt{x^2+y^2}$.
For example, for a vector $T^i$, $T^i(x,y,z) = {R^i}_j T^j(\rho, 0, z)$.
Equation~\eqref{basicrot} describes how to compute the values of a tensor
at points outside the half-plane $x \geq 0$ and $y \neq 0$ from points
within this plane.

One type of non-tensor that is often used in numerical relativity is
the partial derivative of tensors,
$\partial_k{T^{i_1,i_2,\ldots}}_{j_1, j_2,\ldots}$. Under coordinate
transformations, the index $k$ transforms by the chain rule of partial
derivatives with the same Jacobian factor as appears in the coordinate
transformation of a tensor,~\eqref{tensor-coord-xform}.
Using~\eqref{tensor-coord-xform} and the product rule, there appear additional
terms containing $\partial_k{R^i}_j$. However, using coordinates that
are adapted to the axisymmetry~\eqref{rotation-coord-xform} implies
$\partial_k{R^i}_j=0$ (in other words, $R$ describes a rigid
rotation), and hence the basic transformation law~\eqref{basicrot} is
also valid for partial derivatives of tensors in this context.


\subsection{1D interpolation in the $x$~direction}
\label{sec:interpolation}

We now turn to a brief discussion of the necessary interpolations.  As
made explicit in~\eqref{basicrot}, we require fields at points
$(\sqrt{x^2+y^2},0,z)$. As such points are not necessarily part of the
numerical grid, a one-dimensional interpolation may be required. In
this work we test Lagrange polynomial interpolation
(\eg{}~\cite{FMM77,Press86}) with interpolation polynomials of
degrees~2 through~5, obtaining good results already with second order
polynomials.  Our evolution code uses second order finite difference
molecules of size~$v=w=1$, but for interpolation polynomials of degree
larger than~2 we need $v > 1$. Note that even the points at $x = x_{\max}$,
$y \neq 0$ can be obtained through interpolation by first applying the
physical boundary for $y = 0$, then rotating to $y\neq0$ for $x<x_{\max}$,
and then applying the physical boundary at $x=x_{\max}$, $y\neq0$.

This sort of interpolation is used in many areas of computational
science (\eg{}~in adaptive mesh refinement), and thus should not be
expected to pose a serious threat to the success of this technique.
The principal issue of concern is whether interpolation might
destabilize an evolution scheme that often is carefully crafted using
particular forms of centered stencils and averages.  This must be
addressed on a case-by-case basis, either by Von~Neumann or other
analytical stability analysis, and/or by numerical experiment.

Although we do not do it in practice, instead of explicitly computing
and storing field values at the boundaries in the $y$~direction, one
can make the polynomial interpolation a part of the stencil, thereby
explicitly reducing the 3D stencil to an inhomogeneous and asymmetric --
but nonsingular -- 2D stencil.  In contrast, storing field values for
a 3D stencil allows us to simply use numerical routines from existing
Cartesian 3D codes.


\section{Applications}
\label{sec:apps}

In this section we provide two important tests, both with the complete
set of 3D, nonlinear Einstein evolution equations, performed with
``Cartoon'' in the Cactus code for numerical
relativity~\cite{Cactusweb,Allen99a,Bona98b,Seidel98c}.  However, we stress
that the techniques developed here are applicable to many families of
partial differential equations.  That
our tests are so successful, with such a complicated set of tensor PDEs
as the $3+1$ Einstein equations involving dozens of coupled nonlinear
evolution equations, stresses this point. Simpler sets of equations
can be expected to have fewer complications: less severe regularity issues
at symmetry points
(origin or axis), simpler transformation laws in the rotation to provide
boundary conditions, etc.


\subsection{Einstein's Equations}

In this section we give a brief introduction to the basic equations we
will be solving.  The Einstein equations of general relativity in 3+1
form (see \cite{York79,York83} for recent reviews and further
references) are a complicated set of coupled, nonlinear partial
differential equations for the symmetric tensor fields $\gamma_{ij}$
and $K_{ij}$ (indices range from $1$ through~$3$, so there are
12~field variables in all). The metric $\gamma_{ij}$ is the spatial
part of the spacetime metric (which gives the invariant distance
between two infinitesimally separated events):
\begin{equation}
 ds^2 = -(\alpha^2 -\beta^i\beta_i) \, dt^2 + 2 \beta_i \, dx^i dt
+\gamma_{ij} \, dx^i dx^j.
\end{equation}
The extrinsic curvature, $K_{ij}$, specifies how the $t = \constant$
slices are embedded in spacetime.  The equations also involve the
auxiliary tensor fields $\alpha$ and $\beta^i$. These so-called
``gauge'' fields carry no dynamical
information: they may be freely chosen
for convenience.  The lapse function $\alpha$ determines the proper time
$d\tau =\alpha \, dt$ measured by an observer falling normal to the
time slice defined by $t=\constant$.  The shift vector $\beta^i $ determines
the coordinate distance a constant-coordinate point moves away from the
normal vector to the slice as one advances from one slice to the 
next.  In the tests performed in this paper, we choose $\beta^{i}=0$ for
simplicity.

The evolution equations can be written as:
\begin{mathletters}
						\label{eqn:evolution}
\begin{equation}
\partial_t \gamma_{ij} = -2 \alpha K_{ij} + D_i \beta_j + D_j \beta_i,
\end{equation}
\begin{eqnarray}
\partial_t K_{ij} =&&-D_i D_j \alpha + \alpha [R_{ij} + (\tr K) K_{ij} -
2 K_{ik} K^k{}_j] \nonumber \\
&&+ \beta^k D_k K_{ij} + K_{ik} D_j \beta^k + K_{jk} D_i
\beta^k.
\end{eqnarray}
\end{mathletters}
In these equations $R_{ij}$ is the 3-Ricci tensor, $R$ is the 3-scalar 
curvature (both nonlinear functions of the metric $\gamma_{ij}$ and its
first and second spatial derivatives), $\tr K$ is the trace of $K_{ij}$,
and $D_i$ is the covariant derivative associated with the 3-metric
$\gamma_{ij}$.  With suitable choices for $\alpha$ and $\beta^i$,
equations~\eqref{eqn:evolution} are hyperbolic in $\gamma_{ij}$ and
$K_{ij}$.

The fields $\gamma_{ij}$ and $K_{ij}$ are not completely freely
specifiable: on each $t = \constant$ slice they must satisfy the four
constraint equations
\begin{mathletters}
						\label{eqn:constraints}
\begin{equation}
H \equiv R + (\tr K)^2 - K_{ij} K^{ij} = 0,
					\label{eqn:Hamiltonian-constraint}
\end{equation}
\begin{equation}
H^i \equiv D_j (K^{ij} - g^{ij} \tr K) = 0,
\end{equation}
\end{mathletters}
where for later use we define the left hand side functions as $H$
(known as the Hamiltonian constraint) and $H^i$.
Eqs.~\eqref{eqn:constraints} are elliptic equations in $\gamma_{ij}$
and $K_{ij}$; in general they must be solved numerically in order to
obtain valid initial data (\cite{York-Piran-1982-in-Schild-lectures}).
However, one can show that once the constraints are satisfied on an
initial slice, they are preserved by the evolution
equation~\eqref{eqn:evolution} (\ie{} they stay
satisfied for all future times).  This statement holds for the continuum
equations~\eqref{eqn:evolution} and~\eqref{eqn:constraints}; a finite
difference evolution will in general only approximately satisfy the
constraints at later times (\cite{Choptuik91}), and in fact the deviations
of $H$ and $H^i$ from zero are useful diagnostics of the evolution's numerical
accuracy.

To understand the physical meaning of these equations it is useful to
consider an analogy with Maxwell's equations in a vacuum: $\gamma_{ij}$
and $K_{ij}$ are analogous to the (vector) electric and magnetic fields
$\E$ and $\B$, the evolution equations~\eqref{eqn:evolution} are
analogous to the Maxwell equations $\partial_t \E = \del \times \B$
and $\partial_t B = - \del \times \E$, and the constraint
equations~\eqref{eqn:constraints} are analogous to the Maxwell
equations $\del \cdot \E = \del \cdot \B = 0$, with $H$ and $H^i$
analogous to $\del \cdot \E$ and $\del \cdot \B$.
However, unlike Maxwell's equations, the Einstein equations are
nonlinear, quite complicated (they have on the order of 1000~terms when
written out in scalar form in terms of coordinate partial derivatives),
and second order in space (though still first order in time).  These
properties make the Einstein equations difficult to treat
numerically.

As the main point of our paper is a technique for solving {\em any\/}
set of evolution equations, we simply state that the particular form
of the equations we use for this paper is a variant
of~\eqref{eqn:evolution} and~\eqref{eqn:constraints} recently put
forth by Baumgarte and Shapiro~\cite{Baumgarte99}, based on previous
work of Shibata and Nakamura~\cite{Shibata95}. We will refer to this
as the BSSN formulation. It is to be noted that this formulation is
quite similar in many respects to the Bona-Mass\'o
formulation~\cite{Bona92,Bona94a,Bona97a}. In another
paper~\cite{Alcubierre99d} we detail experiments carried out with
these formulations on various spacetimes.  As these details are not
important to the results presented here, in this paper we focus only
on the technique and its application to a very general class of
partial differential equations, as represented by the Einstein
equations.


\subsection{Black hole spacetimes}
\label{sec:holes}

In this section we report on the application of the Cartoon technique to the
nonlinear evolution of black holes in the strong field regime.
To illustrate the power of this technique, we focus on the case of a
Schwarzschild black hole evolved with geodesic slicing.  Geodesically
sliced black hole evolutions have been used extensively to test black
hole codes in 1D \cite{Bernstein89} and 2D
\cite{Bernstein93b,Camarda97a} in polar-spherical type coordinates,
and in 3D~Cartesian
coordinates~\cite{Anninos94c,Bruegmann96,Camarda97a,Camarda97c}.  This
black hole case has the advantages of (i) having an analytic and also a highly
accurate numerical 1D solution; (ii) providing a demanding
test, as the slicing rapidly approaches the spacetime singularity, and
hence spacetime curvature and metric functions grow rapidly and
without bound until the code crashes at time $t=\pi M$, where $M$ is
the mass of the black hole; (iii) being one of the simpler testbeds in
strong field regimes, as the geodesic slicing condition ($\alpha=1$)
does not involve coupling additional elliptic or parabolic equations,
which could complicate the demonstration of the technique.  In the
following section we consider a more complex system as a further test
of the Cartoon method.

The initial 3-metric for the Schwarzschild black hole is given by
\begin{equation}
\label{initdata}
ds^2 = \psi^4 (dr^2 + r^2(d\theta^2 + \sin^2\theta d\phi^2)),
\end{equation}
where the conformal factor is $\psi = (1+\frac{M}{2r})$.  Here $r$ is
the isotropic radius, related to the standard Schwarzschild radius
$r_s$ by $r_s = (1+\frac{M}{2r})^2 r$. Transforming to Cartesian
coordinates, we have
\begin{equation}
\label{cartmetric}
ds^2 = \psi^4 (dx^2 + dy^2 + dz^2),
\end{equation}
where the Cartesian coordinates $x$, $y$, and $z$ are related to the
isotropic radius $r$ in the usual way.  The extrinsic curvature for 
the time symmetric, $t=0$ slice of this spacetime vanishes. 

We evolve this black hole in Cartesian 
coordinates, using the Cartoon technique.  While typical 3D 
simulations of the Einstein equations are currently limited to grid 
sizes of roughly $100^{3}$ (ignoring the possibility of mesh 
refinements~\cite{Bruegmann97}), and the largest production 
simulations are currently at around 
$300^{3}-400^{3}$~\cite{Allen98a,Seidel99b} 
on a 128Gbyte SGI/Cray Origin~2000 supercomputer with 256 processors, 
in this simulation we compute the evolution on a grid of $N_{x} \times 
N_{y} \times N_{z} = 1025 \times 3 \times 2049$.  (Our current 
implementation of the technique requires the $z$~direction to range 
from $+z_{\max}$ to $-z_{\max}$.)  This is equivalent to a full 3D 
calculation of $2049^{3}$, over two orders of magnitude larger than 
the largest production simulations to date.  Yet with this technique, 
the calculation requires only~0.07\% of the memory required to do the
full 3D simulation, and was run on 8~processors of an Origin~2000 
over a period of 40~hours.

In more detail, the present calculation was performed for a unit
mass black hole, with $\Delta x = \Delta y = \Delta z = 0.01$, utilizing an
iterative-Crank-Nicholson (ICN) method of lines evolution scheme
with three iterations,
a time step $\Delta t = 0.1, spatial grid resolution
\Delta x = 0.001$, and 4th~order Lagrange
polynomial interpolation for the Cartoon algorithm.  The total number
of time steps for such a simulation was 3000 (9000 ICN iterations),
corresponding to a time of $t=3.0$.

\begin{figure}      
\epsfxsize=10cm   
\epsfysize=10cm   
\hspace{3cm}\epsfbox{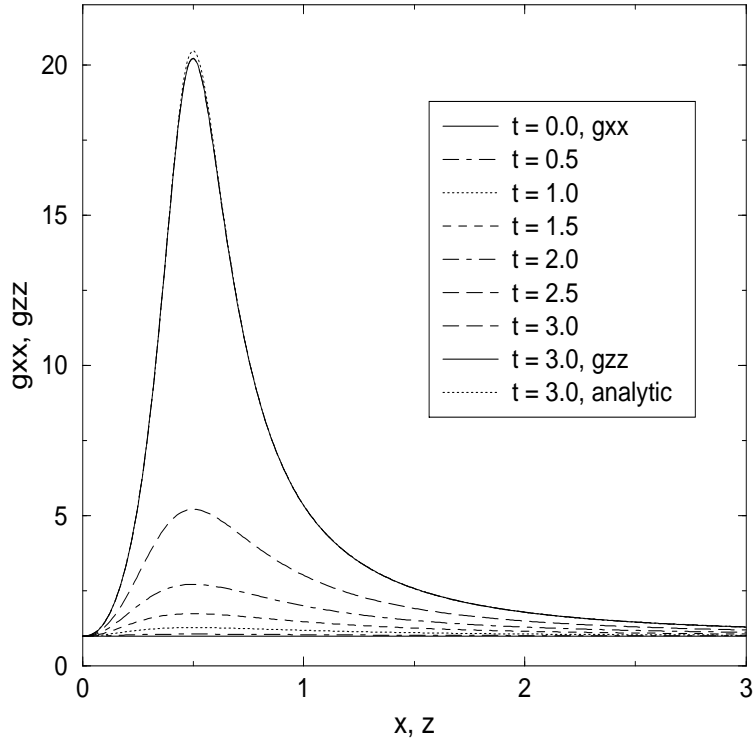}
\caption{
We show results for a geodesically sliced Schwarzschild black
hole, evolved in 3D with the Cartoon technique.  The metric function
$g_{xx}$ is shown on the $x$~axis at times $t/M=0.0, 0.5, \ldots,
3.0$, where $M$ is the mass of the black hole. At $t/M=3.0$, $g_{zz}$
on the $z$~axis is also plotted, but it is indistinguishable from
$g_{xx}$ as expected for spherical symmetry. Both agree well with the
analytic result.  
}
\label{fig:geodesic1}      
\end{figure}

In Fig.~\ref{fig:geodesic1} we show the metric function $g_{xx}$ on
the $x$~axis at selected times during the evolution. There is no hint
of any instability in the result.  At $t = 3.0$, a rather small
difference between the numeric and analytic result 
(cmp.\ \cite{Bruegmann96,Bona98b}) can be seen at the peak of $g_{xx}$. 
At $t = 3.0$ we also show $g_{zz}$ on the $z$~axis, which at this scale is
indistinguishable from the numerical $g_{xx}$ on the $x$~axis, as it
should be for this spherically symmetric data. The $z$~direction is
computed quite differently from the $x$~direction, and suffers from
the maximal inhomogeneity of the Cartoon stencil due to the
interpolation near the $z$~axis. Yet results along the $z$-axis remain
accurate and stable until the end of the simulation, giving a strong
indication of the robustness of the technique.

\begin{figure}      
\epsfxsize=10cm   
\epsfysize=10cm   
\hspace{3cm}\epsfbox{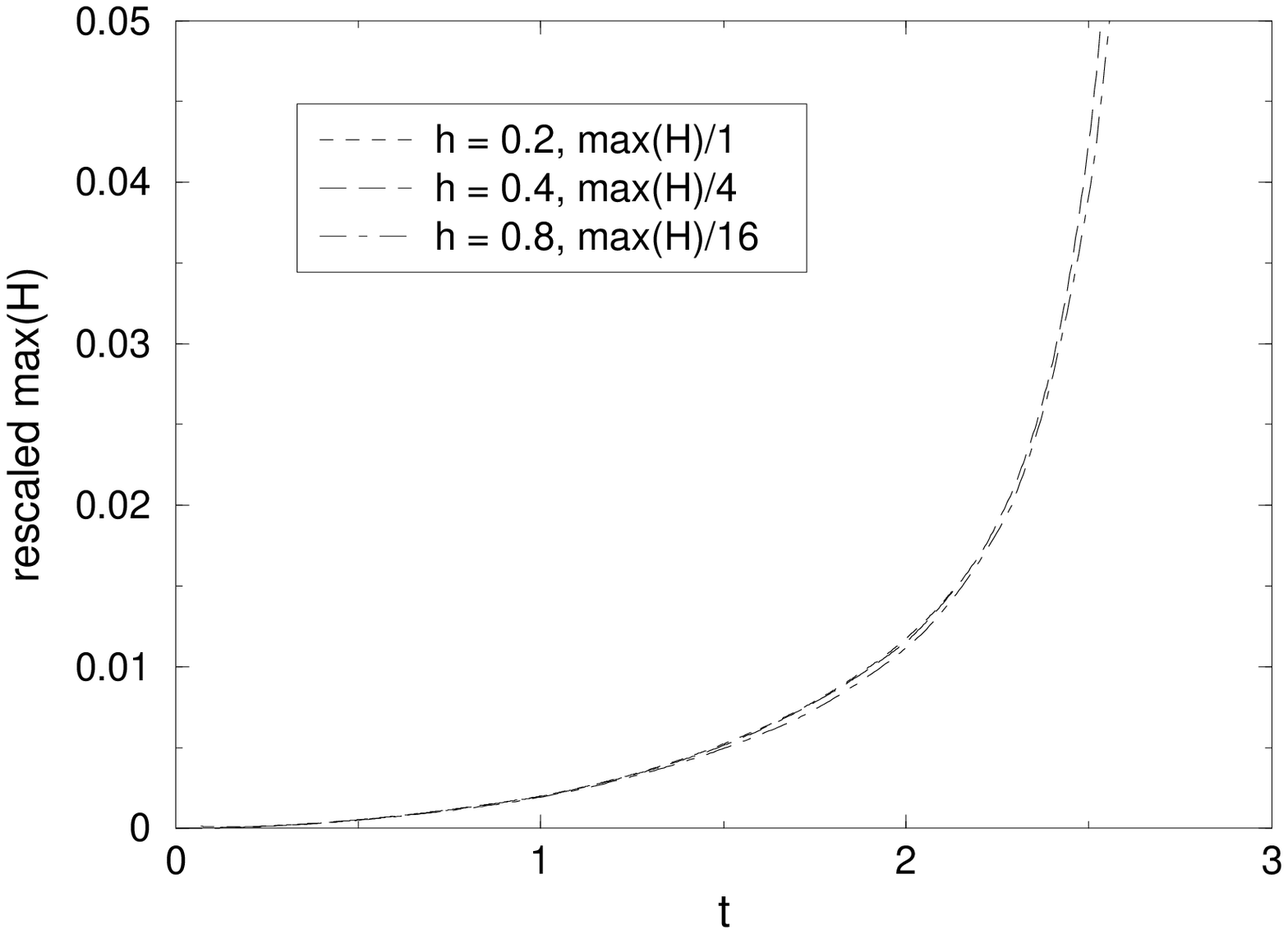}     
\caption{
The maximum of the Hamiltonian constraint for a geodesically sliced
black hole at three different resolutions. Each graph is
scaled proportionally to the
inverse square of the grid resolution.  The equality of the rescaled
values for the different resolutions indicates second order convergence.
}                   
\label{fig:geodesic2}      
\end{figure}

In Fig.~\ref{fig:geodesic2} we demonstrate second order convergence
of the Cartoon evolution.  We use the Hamiltonian constraint~$H$ as a
diagnostic of the evolution's numerical accuracy: since $H$ vanishes
analytically, it should converge to zero as the grid resolution increases.
In particular, for second order finite differencing,
each time the grid resolution is doubled
$H$ should shrink by a factor of~4 (\cite{Choptuik91}).
The figure shows the maximum value
of $H$ in the grid as a function of time, plotted for the 3~different
grid resolutions $\Delta x = \Delta y = \Delta z = 0.02, 0.04, 0.08$,
with each successive plot divided by $1, 4, 16$ respectively, so the values
should be identical at the different resolutions.  Second order
convergence is directly evident from the figure.

It is to be noted that not only is such a simulation impossible today 
in full 3D due to memory constraints, but the 2D Cartoon run
has far higher resolution 
than has been achieved even in 2D codes designed to treat similar 
problems.  Axisymmetric simulations of black holes, usually performed 
in spherical-polar coordinates, are typically not performed with more 
than 300 radial and 50 angular 
zones~\cite{Brandt94b,Brandt94c,Anninos94a,Anninos94b}.  Inherent axis 
instabilities usually create difficulties in simulations with higher 
resolution than this, in part because angular zones closer to the axis 
have more delicate regularity behavior to enforce.
It is precisely these sorts of problems Cartoon is designed to overcome.

The problem with standard axisymmetric codes are exacerbated unless
certain gauges are used that force troublesome terms in the 
three--metric to vanish.  For example, as discussed in 
Ref.~\cite{Bernstein93b}, a vicious axis instability arose in
simulations of spherical or distorted black holes in axisymmetry, 
causing a rapid code crash, until a gauge condition was developed that 
diagonalized the three--metric.  This condition required the solution 
of an elliptic equation to determine $\beta^i$ at each time step, which
is very time consuming to solve numerically.  Furthermore, even with
this gauge choice, if the code is run with too high a resolution,
axis instabilities are again encountered, and these instabilities
eventually crash the code at late times.  Similar 
problems with axis instabilities have been encountered in the case of 
rotating~\cite{Thornburg93,Brandt94a,Brandt94b,Brandt94c} and colliding
black holes~\cite{Cadez71,Smarr75,Eppley75,Smarr79,
Anninos94a,Anninos94b,Anninos97b}.

As reported recently in Ref.~\cite{Camarda97c}, a case similar to the 
one presented here, a geodesically sliced perturbed black hole, was 
studied to compare results from a 3D code in Cartesian coordinates to
a traditional 2D axisymmetric code.  In order to compare metric 
functions directly, the codes had to be run with the same spatial 
gauge, which had vanishing shift.  Although the agreement was 
excellent for as long as it could be computed, due to the axis instability
the axisymmetric code
crashed far earlier than the time at which the slice hit the
singularity.  On the other hand, the 3D 
Cartesian code, due to its lack of coordinate singularities, was able to run 
accurately all the way to the singularity.  With our Cartoon technique, 
we are able to do the same simulation in Cartesian coordinates, with 
the same accuracy and stability, but with far less memory than is 
required in full 3D, and with far more stability than is possible in a 
standard axisymmetric code.


\subsection{Gravitational wave spacetimes}
\label{sec:waves}
While the above simulation of time slices approaching a black hole 
singularity provide a strong test of the techniques we have 
developed, the underlying system is spherically symmetric, and does 
not contain gravitational waves.  In order to further test and confirm this 
technique, we now turn to a completely different spacetime system, 
one not initially containing no black hole.
Instead, the system we choose is that 
of highly nonlinear gravitational waves.  

Gravitational waves are often considered in the linearized regime, 
where they are small disturbances on some background (often flat) spacetime
that propagate at the speed of light.  However, as the 
Einstein equations are nonlinear, for cases where the waves have 
sufficiently large amplitude they can affect the background spacetime 
on which they propagate.  Even more, for strong enough waves there 
is no background spacetime;  the waves have to be treated fully 
nonlinearly, and under extreme conditions they can even collapse in 
on themselves, forming a black hole when none existed previously.

Low amplitude gravitational waves have provided testbeds of 3D 
numerical relativity codes for the last decade, and even there they 
have provided a strong challenge~\cite{Shibata95,Anninos94d,Bona98b}.  
Extreme gravitational wave simulations, where the waves form their 
own background, and where they may actually form black holes, have 
only been possible to evolve in full 3D codes during the last 
year~\cite{Alcubierre99b}.  For waves above a certain 
critical amplitude a black hole forms, while for
waves just below critical a rich pattern of oscillations 
develops as the waves teeter on the edge of forming a black
hole, and then eventually they
disperse and the system returns to flat space (in a highly 
nontrivial coordinate system).

In this paper we use one such simulation to illustrate the 
strength of the Cartoon technique, and compare with a full 3D simulation as 
a measure of its accuracy.  As our main purpose here is to demonstrate 
that the technique is robust, even under very demanding simulations, 
we will not go into detail of the physics of these simulations.  The 
interested reader is asked to consult Ref.~\cite{Alcubierre99b} for 
more details.

As in Ref.~\cite{Alcubierre99b}, we take as initial data a pure 
gravitational wave data set, based on the axisymmetric ansatz of 
Brill~\cite{Brill63}, and later studied by 
Eppley~\cite{Eppley75,Eppley77,Eppley79} and others~\cite{Holz93,Gentle99}.
The metric takes the form
\begin{equation}
ds^2 = \Psi^4 \left[ e^{2q} \left( d\rho^2 + dz^2 \right) 
+ \rho^2 d\phi^2 \right] =\Psi^4 \hat{ds}^{2},
\label{eqn:brillmetric}
\end{equation}
where $q$ is a free function subject to certain boundary conditions.  
We consider a function $q$ of the form
\begin{equation}
q = a \; \rho^2 \; e^{-r^2},
\end{equation}
where $a$ is a constants, and $r^2 \equiv \rho^2 + z^2$ (see
\cite{Holz93} for 2D, and~\cite{Alcubierre98b,Alcubierre99b} for full 3D 
Cartesian coordinates).  We consider 
data with the amplitude $a = 4$ which corresponds to a strong,  
axisymmetric, equatorial plane symmetric gravitational wave. 
We choose the extrinsic curvature to vanish (time symmetric initial 
data).  As shown in Ref.~\cite{Alcubierre99b}, this initial data 
collapses in on itself initially, but after a series of reverberations 
it disperses, leaving flat space in its wake.

Taking this form for~$q$, we solve the Hamiltonian 
constraint equation~\eqref{eqn:Hamiltonian-constraint}
numerically using a multigrid elliptic solver on an $N_{x} 
\times N_{y} \times N_{z} = (129 \times 3 \times 257)$ grid, with 
$\Delta x = \Delta y = \Delta z = 0.04$, and the outer boundary at
$5.12$.  As this axisymmetric initial 
data set is also symmetric about all coordinate planes, it is possible 
to evolve it as a full 3D system in just one octant ($x,y,z$ 
non-negative), as was also done in Ref.~\cite{Alcubierre99b}.  The 
same system is solved numerically in full 3D (with octant symmetry) on 
an $N_{x} \times N_{y} \times N_{z} = (129 \times 129 \times 129)$ 
grid, again with $\Delta x = \Delta y = \Delta z = 0.04$, as a
comparison simulation.  

\begin{figure}      
\epsfxsize=10cm   
\epsfysize=10cm   
\hspace{3cm}\epsfbox{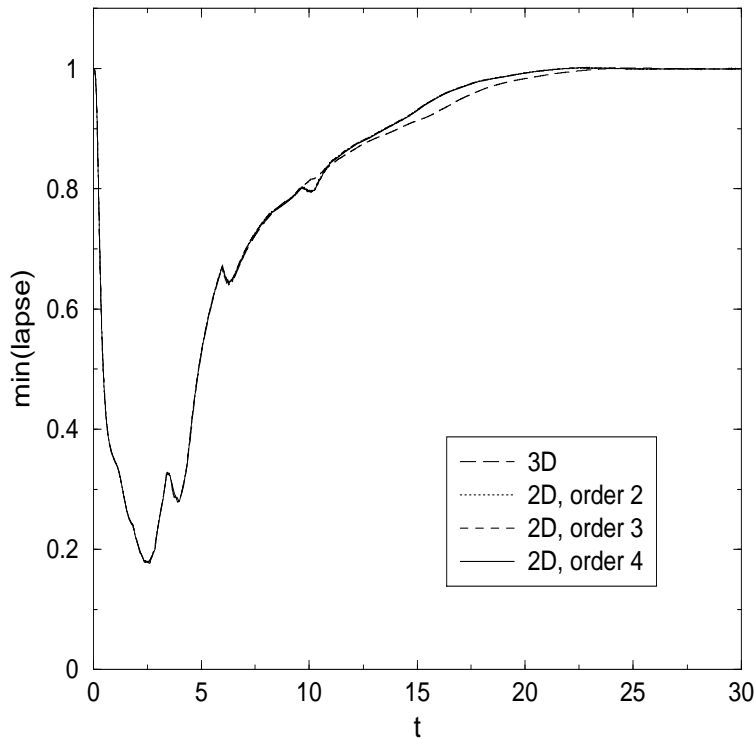}     
\caption{
The minimum of the lapse during the evolution of a strong Brill wave,
which generically occurs at the origin and is a sensitive measure of
the spacetime evolution. We compare three Cartoon runs
using interpolation order 2, 3, and 4, with a 3D run.
}                   
\label{fig:wave1}      
\end{figure}

Both the full 3D system and the slab system were evolved to a time
when the gravitational waves had largely left the system (through
outgoing boundary conditions applied on the evolved function; see
Ref.~\cite{Alcubierre99b} for details).  In Fig.~\ref{fig:wave1}, as a
sensitive measure of the evolution we show the minimum value of the
lapse function $\alpha$ as a function of time throughout the evolution.
We have found this to be a good indicator of the evolution of the
system~\cite{Alcubierre99b}, and also, as it sits on the axis and
origin of the system, it should be especially sensitive to any
problems that arise during the evolution due to the Cartoon procedure.
We show results for 3D, and for Cartoon using interpolation of order 2, 
3, and 4.
The agreement is excellent.  There is a discernible
difference between the 3D and Cartoon runs after time $t=10$, starting with a
slight bump that can be seen in the slab evolution that is not present in
the full 3D simulation.  This is due to slight differences in the
boundary treatment in the two cases.  There is no known perfect
outgoing wave condition for general relativity (nor even a clear way to
identify what a wave is in a nonlinear situation such as this), and
small differences in boundary treatment can lead to differing amounts of
reflection at the boundaries.  The different geometries of the two
simulations lead to different characteristics of the boundary
treatment, which are ultimately reflected in the results. The
origin of this difference is well-understood, to be expected,
and minor, and is
unrelated to the main thrust of this paper.

\section{Conclusions}
\label{sec:concl}

We have developed a novel technique for computational simulations with 
certain symmetries. This Cartoon technique allows simulations
to be performed in 3D~Cartesian 
coordinates, thereby avoiding difficulties associated with coordinate 
singularities that often lead to numerical instabilities.  Compared
to the full 3D~Cartesian approach, Cartoon allows a huge savings
in both memory and computational time, without introducing 
problems commonly associated with the singular coordinate systems which are
usually employed.  The Cartoon technique should be useful for any system of 
partial differential equations, and we have shown that it works well
in one of the most complicated sets of equations in theoretical physics,
Einstein's equations of general relativity.

While this approach to axisymmetric simulation inherits many
of the advantages of 3D~Cartesian simulations, there are also
some disadvantages which carry over. For example, a 2D code using
angular coordinates can align spherical boundaries at constant
coordinate value. Furthermore, a logarithmic or other stretched
radial coordinate can be introduced that significantly improves
resolution for problems with asymptotic $1/r$ fall-off.  Even though
neither of these techniques are readily incorporated in the Cartoon method, we
nonetheless expect Cartoon to have a great impact on numerical relativity,
as general purpose stable 2D codes are not currently available in the field.

Furthermore, as most fully 3D simulations are carried out in the same Cartesian
coordinates as are found in Cartoon,
experience gained from the accelerated Cartoon simulations should carry over
directly to the full 3D work.  This has generally not been the case
previously, as special coordinate systems used for axisymmetric or
spherically symmetric systems also required special treatments and
gauge conditions that simply were not applicable in 3D~Cartesian coordinates.
Hence with our new technique systems can be studied with the
{\em same\/} coordinate systems, with the {\em same\/} gauges, and with
the {\em same\/} analysis tools as they will be when performed in full
3D. Finally, this technique has the potential to allow for a number of
intrinsically axisymmetric systems to be studied with unprecedented
stability and accuracy, since simulations with resolutions of
several thousand grid points on a side are readily achievable.

We have implemented this technique in the Cactus code for 3D numerical 
relativity and the Cactus Computational Toolkit, and we expect that it 
will be a powerful addition to the field of numerical relativity.

{\bf Acknowledgments.} 
The basic idea for this work was suggested by Steven Brandt. Special
thanks are due to Paul Walker for early input on this idea. This work
was supported by the AEI, NSF PHY 98-00973, and FWF P12754-PHY.  The
calculations were performed at the AEI.  We thank many colleagues at
the AEI, NCSA, Univeristat de les Illes Balears, and Washington University
for the co-development of the Cactus code.



\end{document}